\documentstyle [12pt,epsfig] {article}

\newcommand{\beq}{\begin{equation}}
\newcommand{\eeq}{\end{equation}}
\newcommand{\beqar}{\begin{eqnarray}}
\newcommand{\eeqar}{\end{eqnarray}}
\newcommand {\nono} {\nonumber \\}

\newcommand {\bear} [1] {\begin {array} {#1}}
\newcommand {\ear} {\end {array}}

\newcommand {\eqr} [1]
	{(eq. \ref {eq:#1})}

\newcommand {\myI} [1] {\int \! #1 \,}

\newcommand {\beqarn} {\begin{eqnarray*}}
\newcommand {\eeqarn} {\end{eqnarray*}}

\newcommand {\sepe} {\;\;\;\;\;\;\;\;}

\newcommand {\brac} [1]	{{\left\{	#1 \right\}}}
\newcommand	{\paren} [1] {{\left( #1 \right)}}
\newcommand	{\brak} [1] {{\left[ #1 \right]}}

\newcommand {\mrm} [1] {\mathrm {#1}}

\newcommand {\comm} [2] {\mbox {$\left[ #1, #2 \right]$}}

\newcommand {\myref} [1]	%
	{%
	\begin{thebibliography} {99}	%
			{#1}	%
	\end {thebibliography}}

\def\emph#1{{\em #1}}
\def\mathbf#1{{\bf #1}}
\def\mathrm#1{{\rm #1}}
\def\mathit#1{{\it #1}}

\newcommand {\pa}	{{\partial}}

\newcommand {\half}	{\frac 1 2}

\catcode`\@=11
\@addtoreset{footnote}{section}
\catcode`\@=12

\newcounter	{exinsert}	[subsection]

\renewcommand {\theexinsert}	%
{	\thesubsection.\arabic{equation}}

\renewcommand {\theequation} {\thesection.\arabic{equation}}

\catcode`\@=11
\@addtoreset{equation}{section}
\catcode`\@=12

\newenvironment {exinsert} [1]	%
{	
	\begin {quotation}	%
	\refstepcounter {equation}	%
	{\bf {#1}} \theequation. \,\,	%
}
{	
	\end {quotation}
}

\newcommand {\bprop}
{\begin {exinsert} {Proposition}
}
\newcommand {\eprop} {\end {exinsert} }

\newcommand {\bexe} {\begin {exinsert} {Exercise}}
\newcommand {\eexe} {\end {exinsert} }

\newcommand {\bexa} {\begin {exinsert} {Example}}
\newcommand {\eexa} {\end {exinsert} }

\newcommand {\literal} [1] {{ \mbox {$\mrm {#1}$}}}

\newcommand {\grad} {\nabla}

\newcommand {\ncmd} [2] {\newcommand {#1} {#2}}

\ncmd {\vgrad} {{\vec \grad}}

\ncmd {\Mv} {{\cal M}_V}
\ncmd {\Mh} {{\cal M}_H}

\ncmd {\adj} {\literal {adj}}	
\ncmd {\Mp} {M_\literal {planck}}	

\ncmd {\x} {$\times$}

\newcommand {\btab} [3]
  {\begin{table} [htb]
   \begin{center}
   \caption {#2}	\label {tab:#1}
   \begin {tabular} {#3}
   \hline}

\ncmd {\etab}	{   \hline
   \end {tabular}
   \end{center}
  \end{table}}

\def\jou#1,#2,#3,{{\sl #1\/ }{\bf #2} (19#3)\ }
\def\ibid#1,#2,{{\sl ibid.\/ }{\bf #1} (19#2)\ }
\def\am#1,#2,{{\sl Acta. Math.\/ } {\bf #1} (19#2)\ }
\def\annp#1,#2,{{\sl Ann.\ Physics\/ }{\bf #1} (19#2)\ }
\def\cmp#1,#2,{{\sl Comm.\ Math.\ Phys.\/ }{\bf #1} (19#2)\ }
\def\cqg#1,#2,{{\sl Class.\ Quantum Grav.\/ }{\bf #1} (19#2)\ }
\def\dm#1,#2,{{\sl Duke\ Math.\ J.\/ }{\bf #1} (19#2)\ }
\def\ijmpa#1,#2,{{\sl Int.\ J.\ Mod.\ Phys.\/ }{\bf A#1} (19#2)\ }
\def\invm#1,#2,{{\sl Invent.\ Math.\/ }{\bf #1} (19#2)\ }
\def\jhep#1,#2,{{\sl JHEP\/ }{\bf #1} (19#2)\ }
\def\jmp#1,#2,{{\sl J.\ Geom.\ Phys.\/ }{\bf #1} (19#2)\ }
\def\jams#1,#2,{{\sl J.\ Diff.\ Geom.\/ }{\bf #1} (19#2)\ }
\def\jdg#1,#2,{{\sl J.\ Amer.\ Math.\ Soc.\/ }{\bf #1} (19#2)\ }
\def\jpa#1,#2,{{\sl J.\ Phys.\ A.\/ }{\bf #1} (19#2)\ }
\def\grg#1,#2,{{\sl Gen.\ Rel.\ Grav.\/ }{\bf #1} (19#2)\ }
\def\mpla#1,#2,{{\sl Mod.\ Phys.\ Lett.\/ }{\bf A#1} (19#2)\ }
\def\ma#1,#2,{{\sl Math.\ Ann.\/ } {\bf #1} (19#2)\ }
\def\nc#1,#2,{{\sl Nuovo Cim.\/ }{\bf #1} (19#2)\ }
\def\npb#1,#2,{{\sl Nucl.\ Phys.\/ }{\bf B#1} (19#2)\ }
\def\plb#1,#2,{{\sl Phys.\ Lett.\/ }{\bf #1B} (19#2)\ }
\def\pla#1,#2,{{\sl Phys.\ Lett.\/ }{\bf #1A} (19#2)\ }
\def\pnas#1,#2,{{\sl Proc.Nat.Acad.Sci.\/ }{\bf #1} (19#2)\ }
\def\prev#1,#2,{{\sl Phys.\ Rev.\/ }{\bf #1} (19#2)\ }
\def\prd#1,#2,{{\sl Phys.\ Rev.\/ }{\bf D#1} (19#2)\ }
\def\prl#1,#2,{{\sl Phys.\ Rev.\ Lett.\/ }{\bf #1} (19#2)\ }
\def\prpt#1,#2,{{\sl Phys.\ Rept.\/ }{\bf #1C} (19#2)\ }
\def\ptp#1,#2,{{\sl Prog.\ Theor.\ Phys.\/ }{\bf #1} (19#2)\ }
\def\ptpsup#1,#2,{{\sl Prog.\ Theor.\ Phys.\/ Suppl.\/ }{\bf #1} (19#2)\ }
\def\rmp#1,#2,{{\sl Rev.\ Mod.\ Phys.\/ }{\bf #1} (19#2)\ }
\def\sm#1,#2,{{\sl Selec. Math.\/ }{\bf #1} (19#2)\ }
\def\yadfiz#1,#2,#3[#4,#5]{{\sl Yad.\ Fiz.\/ }{\bf #1} (19#2) #3%
\ [{\sl Sov.\ J.\ Nucl.\ Phys.\/ }{\bf #4} (19#2) #5]}
\def\zh#1,#2,#3[#4,#5]{{\sl Zh.\ Exp.\ Theor.\ Fiz.\/ }{\bf #1} (19#2) #3%
\ [{\sl Sov.\ Phys.\ JETP\/ }{\bf #4} (19#2) #5]}

\newcommand{\cH}{{\cal H}}

\newcommand{\ToCite} [1] {\cite {#1}}

\newcommand {\bra} [1] {{\left< #1 \right|}}
\newcommand {\ket} [1] {{\left| #1 \right>}}

\renewcommand {\eqr} [1]
	{(eq. \ref {#1})}

\begin {document}

\begin{titlepage}

\begin{center}
hep-th/9908152 \hfill IASSNS-HEP-99/77 \\
		\hfill August, 1999 \\
\vskip 1 cm
{\LARGE \bf  A Note On Space Noncommutativity
\\}
\vskip 1 cm
{Zheng Yin\footnote {email:  yin@ias.edu}}\\
\vskip 0.75cm
{\sl 	 School of Natural Sciences \\
	Institute for Advanced Study\\
	Olden Lane, Princeton, NJ 08540\\
}
\end{center}
\vskip 0.5 cm
\begin{abstract}
We consider a two-point spatial lattice approximation to an
open string moving in a flat background with $B$
field.  It gives a constrained dipole system under the
influence of a vector potential.
Solving and quantizing this system recover
all the essential features of a noncommutative
space.  In particular, open string
interactions induce a canonical product structure on
the Hilbert space of the dipole system.  It coincides
with the usual star product, even though the position
operators can be thought of as mutually commuting.
Modification of gauge transformations in this noncommutative
space also naturally emerges.
\end{abstract}

\end{titlepage}

\section {Introduction and ideology}

	In the passage from classical to quantum mechanics,
the physical observables of a system, which are functions on
its phase space, become the operators on the Hilbert space of
physical states.  The canonical commutation relations
of Heisenberg
\beq
	\comm {X^\mu} {P_\nu} = \imath \hbar \delta^\mu_\nu
\eeq
require as well as
epitomize this fundamental change.  Thus in the quantum world
the phase space ceases to be a normal ``space,'' because the notion of
it as being made up of well defined ``points'' becomes invalid
 (the uncertainty
principle).  Rather than a function that evaluates at each point
of phase space, a physical observable $W(X, P)$ 
must be understood in its entirety as an operator.
This is how algebraic operations on them are
performed.  In particular, multiplications among operators are
associative but generically noncommutative.  This is
an instance of \emph {noncommutative geometry}.

	By contrast, other parts of the canonical commutation
relations
\beq
	\comm {X^\mu} {X^\nu} = 0
\eeq
indicate that the coordinates on the configuration space form a
set of commuting observable and remain simultaneously
measurable.  The usual point-based geometry of the configuration
space therefore survives.  Wavefunctions, or more generally
fields on it, obtain values at each point and algebraic operations
are performed pointwise.

	It is perhaps only natural
to go one step further and pass onto some noncommutative
geometry for the configuration space as well.  For curiosity's
sake it intrigues to see
what transpires when wavefunctions and fields become
noncommutative objects, but from what sort of physical system would
such circumstances arise in the first place?
One class of such scenarios have recently been uncovered:
the matrix theory \ToCite {BFSS}
naturally accommodates noncommutative tori \ToCite {CDSH}.
Subsequently it is
shown how Heisenberg-like commutation relations such as
\beq
	\comm {X^\mu} {X^\nu} = \imath \Omega^{\mu\nu}
\eeq
apply
to the coordinate fields on the ends of an open string.
In particular, $\Omega$ is nonvanishing if so is the
background for the anti-symmetric rank two tensor $B_{\mu\nu}$
\ToCite {HoChu}.
A noncommutative product of functions of the $X$'s
also appear in the
operator product expansion on the open string boundary
\ToCite {Schomerus,SW}.

	This is of course not the first time
noncommutative geometry has been
advocated to play a role in
(open) string theory.  In \ToCite {WittenNoncomm},
a noncommutative but associative algebra
quite naturally emerged from Type I string theory
and was essential in formulating a string field theory.
However, this underlying ``space'' of that algebra
is very large and
complicated.  It is the
space of open paths in space-time.  Using the
terminology for open string quantization,
it includes not only the wavefunction of the
``center of mass'' coordinates,
but also those of the infinite tower of oscillators.
Somewhere among them one would like 
to discern the most basic features 
of the space-time.  The aim of this brief note is
precisely to extract this piece
of information.  In section two we truncate the theory of
an open string ending on a D-brane to a constrained quantum
mechanical model for its two end points\footnote {
See also \ToCite {BS}.}.
This dipole system inherits a noncommutative product structure
on its Hilbert space from
the open string interaction.
We show that it coincides with the ``star'' product
in deformation quantization in section three.  This model allows
us also to study how the usual gauge transformations
are deformed in noncommutative geometry. Finally, we give some remarks.

\section {The constrained dipole model}

	In the conformal gauge, the free action for an
open string  starting and ending on the same
(single) D-brane is

\beqar
	S &=& \frac 1 {4 \pi \alpha'}
	\int_{\Sigma} \! d^2 \sigma \,
	   \brak { G_{\mu\nu} \paren{{\dot Y}^\mu {\dot Y}^\nu
			- { Y'}^\mu {Y'}^\nu }
		+ B_{\mu\nu} \paren{{\dot Y}^\mu {Y'}^\nu
			- { Y'}^\mu   {\dot Y}^\nu } }	\nono
	&& +  \int_{\pa_2 \Sigma} \! d\tau \,
		A_\mu(Y) \dot Y^\mu
	-  \int_{\pa_1 \Sigma} \! d\tau \,
		A_\mu(Y) \dot Y^\mu.
\eeqar
Here the subscripts ``$2$'' and ``$1$'' on $\pa \Sigma$ 
label the two boundaries of the open string worldsheet.
$G$ and $B$ are constant backgrounds for the metric and rank two
anti-symmetric tensor fields respectively.  $A$ is the background
gauge potential for a constant $U(1)$ field strength on the D-brane.
The term involving $B$ is a locally exact, can be replaced by an
integral on the boundaries, and absorbed by
$A_\mu \to A_\mu + B_{\mu\nu} X^\nu$.
We will do so in the rest of the paper and henceforth drop any explicit
reference to $B$, writing the action more succinctly as
\beqar
	\label {OpenStringAction}
	S &=& \frac 1 {4 \pi \alpha'}
	\int_{\Sigma} \! d^2 \sigma \,
	   \paren{{\dot Y}^2
			- { Y'}^2 }	\nono
	& & 	 +  \int_{\pa_2 \Sigma} \! d\tau \,
		A_\mu(Y) \dot Y^\mu
	-  \int_{\pa_2 \Sigma} \! d\tau \,
		A_\mu(Y) \dot Y^\mu.
\eeqar
Unless otherwise specified, we will also restrict
our attention to the spatial part of the
target space-time for which $F = dA$ is nondegenerate,
assuming that the background allows such a
decomposition.  The full solution of the equation of motion requires
the boundary condition
\beq	\label {OpenStringConstraint}
	\frac 1 {2 \pi \alpha'} G_{\mu\nu} {Y'}^\nu
		= F_{\mu \nu} {\dot Y}^\nu.
\eeq

	Now we approximate the spatial extent of the
open string by a ``lattice'' of two points, situated at the
two boundaries and labeled accordingly.  Let the width of
the string be $2 / \omega$.  Then spatial derivative
$Y'$ is approximated by $\frac \omega 2 ( Y_2 - Y_1 ) $, and the
action \eqr {OpenStringAction} by
\beqar
	S &=&
	\myI {d\tau}
	   \brak { \frac 1 {4 \pi \omega \alpha'}
	\paren{ {\dot Y}_1^2 + {\dot Y_2}^2
			- \frac {\omega^2} 2 ( Y_2 - Y_1 )^2 }}	\nono
	&& +  \myI {d\tau}
		\paren {A_\mu(Y_2) \dot Y_2^\mu
			-  A_\mu(Y_1) \dot Y_1^\mu}.
\eeqar
With parameters
that are more suggestive for particle mechanics,
\beq
	m = \frac 1  {\pi \omega \alpha'}, \sepe
	k = \frac \omega {4 \pi \alpha'}
\eeq
the action now reads as\footnote
	{As usual we use $G$ and $G^{-1}$ to
	lower, raise, and contract indices.}
\beqar	\label {DipoleAction}
	S &=&
	\myI {d\tau}
	   \brak { \frac m {4}
	\paren{ {\dot Y}_1^2 + {\dot Y}_2^2}
			- \frac k 2 ( Y_2 - Y_1 )^2 }	\nono
	&& +  \myI {d\tau}
	\paren {A_\mu(Y_2) \dot Y_2^\mu -  A_\mu(Y_1) \dot Y_1^\mu}.
\eeqar
It is the action of a dipole of total mass $m$ equally divided among 
two points of charge $\pm 1$ linked
by a spring with Hooke constant $k$.  However, one also has to
remember to impose the lattice version of the constraints
\eqr {OpenStringConstraint}
\beqar	\label {DipoleConstraint}
	C_1^\mu &\equiv& (Y_1 - Y_2)^\mu
	   +  \frac 1 k G^{\mu\rho} F_{\rho \nu} \dot Y_1^\nu \sim 0;
\nono
	C_2^\mu &\equiv& (Y_2 - Y_1)^\mu
	   -  \frac 1 k G^{\mu\rho} F_{\rho\nu} \dot Y_2^\nu \sim 0.
\eeqar

	There is no secondary constraint.  Using
\beq
	\brac {Y_2^\mu, \Pi^2_{ \nu}} = \delta^\mu_\nu, \sepe
	\brac {Y_1^\mu, \Pi^1_{\nu}} = \delta^\mu_\nu
\eeq
and
\beqar
	\Pi^2_{\nu}&=& \frac m 2 G_{\nu\mu} {\dot Y}_{2}^{\mu} + A_\nu (Y_2), \nono
	\Pi^1_{\nu}&=& \frac m 2 G_{\nu\mu} {\dot Y}_{1}^{\mu} - A_\nu (Y_1)
\eeqar
we find that
the Poisson brackets of the $C$'s are
\beqar
	\brac {C_2^\mu, C_2^\nu} &=&
		\frac 4 {m k} F^{\mu\mu'} \paren {\delta_{\mu'}^\nu
		- \frac 1 {mk} F_{\mu'\nu'} F^{\nu' \nu}}
	\equiv T^{\mu\nu}, \nono
	\brac {C_1^\mu, C_1^\nu} &=& - T^{\mu\nu}, \nono
	\brac {C_1^\mu, C_2^\nu} &=& 0.
\eeqar
We also find
\beq
	\brac {Y_2^\mu, C_2^\nu} = \frac 2 {mk} F^{\mu\nu}, \sepe
	\brac {Y_1^\mu, C_1^\nu} = - \frac 2 {mk} F^{\mu\nu}
\eeq
Now following the standard procedure for
solving constraint on phase space,
we compute the Dirac brackets
\beqar	\label {CommutatorOnBoundary}
	\brac {Y_2^\mu, Y_2^\nu}_* &\equiv&
	\brac {Y_2^\mu, Y_2^\nu}
	- \brac {Y_2^\mu, C_2^{\mu'}} T^{-1}_{{\mu'}{\nu'}}
		\brac {C_2^{\nu'}, Y_2^\nu}	\nono
	&=& \frac 4 {m^2k^2} F^{\mu\rho}
		T^{-1}_{\rho\rho'} F^{\rho'\nu}	\nono
	& = &  (2 \pi \alpha')^2 F^{\mu\mu'}
	\paren {\delta_{\mu'}^\nu - (2 \pi \alpha')^2
		F_{\mu'\nu'}F^{\nu'\nu}}	\nono
	&\equiv& \Omega^{\mu\nu},	\nono
	\brac {Y_1^\mu, Y_1^\nu}_* &=& - \Omega^{\mu\nu}, \nono
	\brac {Y_2^\mu, Y_1^\nu}_* &=& 0.
\eeqar
These are precisely the Dirac brackets for coordinate fields
on the string boundaries found in \ToCite {HoChu}.

	It turns out to be instructive to use the
variables
\beq
	X^\mu = (Y_2^\mu + Y_1^\mu) / 2, \sepe
	S^\mu = (Y_2^\mu - Y_1^\mu) / 2.
\eeq
In term of them, the quantum commutation relations for the constrained
system are
\beq	\label {xsCoordinates}
	\comm {X^\mu} {S^\nu} = \imath \half {\Omega^{\mu\nu}}
\eeq
but at the same time
\beq
	\comm {X^\mu} {X^\nu} = 0 = \comm {S^\mu} {S^\nu}
\eeq
Therefore the $X$'s and the $S$'s separately each make a
complete set of commuting observables.  Since the $X$'s are
the center of mass coordinates of the dipole system,
that they commute seems at first sight
contrary to the spirit of noncommutative geometry,
Yet we shall show presently that it is precisely in terms of these
\emph {commuting} coordinates
that a familiar type of noncommutative algebra naturally emerges.

\section {Space Noncommutativity}

	In \ToCite {WittenNoncomm} it is observed that
Type I string theory naturally
gives rise to a noncommutative product:  
gluing the right end of one open string to the left end of another
yields a new open string.  This process is at once obviously
noncommutative but associative.  It is also much in the spirit of
the Chan-Paton factors in resembling the multiplication of
matrices through the contraction of adjacent covariant and contravariant
indices:
\[
	(M N)^a_b \equiv M^a_c N^c_b.
\]
Open string fields, however, are big and unwieldy
objects, and there are varying schemes to implement the actual
product.  However, any good product should have one thing in common.
It should be compatible with the way open strings interact, just
as in field theory pointwise multiplication of fields (wavefunctions)
represent their pointwise interaction in a second quantized theory.
All purely open string interactions can be reduced to the process of
two open strings joining together at one boundary and merge into another
open string.  This is what a good product of open string fields 
should describe.  What happens to the interior
of the strings must be complex and one needs to perform 
gauge fixing to describe it, but at least the wavefunctions
of two ``adjacent''
end points  get ``contracted.''
On the other hand, in reducing the
open string to the dipole system \eqr {DipoleAction},
the end points are what is left of the string.
Thus string interaction induces a canonical
product on the Hilbert space $\cH$
 of the dipole system.

	Because the $X$'s form a complete set of
commuting observables, states in $\cH$
space can be written as functions of them.  To write out the sought
product, however, it is convenient to choose another basis.
In light of \eqr {CommutatorOnBoundary}, $\cH$ has a tensor product
structure  $\cH_2 \otimes \cH_1$.  $\cH_i$ is the Hilbert space
associated with the i-th endpoint, furnishing a representation for $Y_i$
in \eqr {CommutatorOnBoundary}.  $\cH_2$ and $\cH_1$ are canonically
isomorphic.  Choosing a basis of states $\ket {z}_i$
for $\cH_i$, the product on $\cH$ is defined by
\beq \label {NoncommutativeProduct}
	\paren {\ket {z_2}_2 \otimes \ket {z_1}_1 }
	* \paren {\ket {z'_2}_2 \otimes \ket {z'_1}_1}
		\to
	 (\bra {z_1} 	 \ket {z'_2} )
	\ket {z_2}_2 \otimes \ket {z'_1}_1.
\eeq

	The explicit calculation is carried out by first choosing from
the
$Y_2$'s a maximal set of commuting operators
$\brac{Z_2}$.  The corresponding
operator $\brac{Z_1}$  from the $Z_1$'s
is similarly a maximal commuting set in light of
\eqr {CommutatorOnBoundary}.  We also divide the $X$'s into
$\hat X = (Z_1 + Z_2) / 2$ and $\tilde X$ being the rest.
The eigenstates of the $Z_i$'s
furnish a set of basis for $\cH_i$ suitable for use in
\eqr {NoncommutativeProduct}.
Using \eqr {CommutatorOnBoundary}, one find the conversion matrix
elements between this and the $x$ basis that diagonalizes the $X$'s:
\beq
	\bra {x} \paren { \ket {z_2}_2 \otimes \ket {z_1}_1}
	= \delta \paren {\hat x -  (z_2 + z_1) / 2}
	\exp{\paren {\frac \imath 4 \Omega^{-1}_{\mu\nu}
		\tilde x^\mu (z_2^\nu - z_1^\nu)}}.
\eeq

	Using this, we find that the product in
\eqr {NoncommutativeProduct} is explicitly given by
\beq	\label {StarProduct}
	(\psi_1 * \psi_2) [x] = \left. \exp {\paren {\frac \imath 2
		\Omega^{\mu\nu} \frac \pa {\pa{x'^{\mu}}}
		 \frac \pa {\pa{x''^{\nu}} }}}
	\psi_1(x') \psi_2(x'') \right|^{x'' = x' = x}.
\eeq
It coincides with the star product studied by mathematicians
in deformation quantization.

	This product structure also implies two maps from a state $\psi(x)$ in
 $\cH$ to an operator $\psi_*$
acting on $\cH$.  These maps are one-to-one but not onto.
To see what kind of operators one can obtain from a wavefunction
$\psi(x)$, note that by \eqr {xsCoordinates} $S$ is realized as
an operator on $\cH$ as follows:
\beq
	S = -\frac \imath 2 \Omega^{\mu\nu} \frac \pa {\pa {x^\nu}}
\eeq

Therefore we can rewrite \eqr {StarProduct} as\footnote
	{On account that
	$\Omega$ is anti-symmetric, there is no operator ordering
	ambiguity in treating $\psi_1(X+S)$, $\psi_2 (Y_2)$, or the like, 
	as formal power series.}
\beq
	(\psi_1 * \psi_2) [x] = \psi_1 (X + S) \psi_2(x)
	\equiv \psi_{1 *} \psi_2 (x)
\eeq
with
\beq
	\psi_* \equiv \psi (Y_2) = \psi (X+S).
\eeq
being understood as an operator.
It is easy to convince oneself that this map is an algebra homomorphism:
\beq
	\psi_{1*} \psi_{2*} = (\psi_1 * \psi_2)_*.
\eeq
Treating \eqr  {StarProduct} as an operator $\psi_{*'}$
acting on the state $\psi_1$ gives us an analogous map
\beq
	\psi_{*'} \equiv \psi (Y_1) = \psi (X-S)
\eeq
with the property
\beq
	\psi_{1*'} \psi_{2*'} = (\psi_2 * \psi_1)_{*'}.
\eeq
In more abstract language, $\psi_{1*}$ and $\psi_{2*'}$ act on the
same module $\cH$ from left and right respectively.
They always commute.

	We can now deduce how gauge transformation would
affect the constrained dipole.  The action \eqr {DipoleAction}
by itself implies that under an infinitesimal transformation
$A_\mu \to A_\mu + \pa_\mu \alpha (x)$, the wavefunction
changes its phase by a position dependent
$\exp^{\imath \alpha (Y_2) - \imath \alpha (Y_1)}$.
The constraints \eqr {DipoleConstraint} change the story by
making the $Y$'s noncommuting operators.  Under an infinitesimal
$\alpha$, the wavefunction changes by
\beq
	\delta \psi (x) = i (\alpha_* - \alpha_{*'}) \psi (x)
	= i (\alpha (x) * \psi (x) - \psi (x) * \alpha (x)).
\eeq
For finite $\alpha$,
\beq
	\psi (x) \to \exp^{*\imath \alpha(x)} *
	  \psi(x) * \exp^{* (- \imath \alpha(x))}
	= \exp^{\imath \alpha_*} \exp^{- \imath \alpha_{*'}} \psi (x)
\eeq
where
\beq
	\exp^{* f(x)} \equiv \sum_{n=0}^\infty
	  \frac 1 {n!} f(x)*f(x)* \cdots \mbox {n times} \cdots * f(x).
\eeq
The originally Abelian $U(1)$ gauge group now becomes non-Abelian.

\section {Remarks}

	After the work reported here had been
completed, 
a preprint 
appeared \ToCite {BS} that discussed some related issues and
employed a quantum mechanical model
resembling the one used here but without the constraints
\eqr {DipoleConstraint}.

	Subsequently, 
another preprint on related subjects appeared 
\ToCite {SW}\footnote
{  Aspects of their work had been reported by those authors
earlier at the U. Penn. conference in May as well as String 99.}.
Among other issues, it studied
gauge bosons on
D-branes with constant background B field in relation
to noncommutative geometry.  By contrast, gauge boson dynamics 
is not visible in the model studied in the present paper.   
This is because in throwing out
the bulk degrees of freedom to obtain \eqr {DipoleAction}
one removes all the oscillator parts of the string field.  Hence
only a  scalar particle, namely the tachyon,
remains  in \eqr {DipoleAction}.
One can try to include the photon with a more elaborate model.  
On the other hand, while its physics is relatively simple, 
\eqr {DipoleAction} can serve as a systematic  and tractable ``probe''
for noncommutative ``space'' in a way akin to how D0-branes can
probe various objects in string theory.

\section* {Acknowledgment}
	We would like to thank P.~Aschieri, K.~Bardakci, B.~Morariu,
H.~Ooguri, N.~Seiberg and E.~Witten
for useful discussion.  The research reported here
is supported by NSF grant PHY-9513835.
Most of it was done while the author was visiting
the particle theory group at Lawrence Berkeley Laboratory.
Their hospitality is  hereby gratefully acknowledged.

\myref {
\bibitem{BFSS}
T.~Banks, W.~Fischler, S.H.~Shenker and L.~Susskind,
``M theory as a matrix model: A Conjecture,''
Phys.\ Rev.\ {\bf D55}, 5112 (1997)
hep-th/9610043.

\bibitem{CDSH}
A.~Connes, M.R.~Douglas and A.~Schwarz,
``Noncommutative geometry and matrix theory: Compactification on tori,''
JHEP {\bf 02}, 003 (1998)
hep-th/9711162.
M.R.~Douglas and C.~Hull,
``D-branes and the noncommutative torus,''
JHEP {\bf 02}, 008 (1998)
hep-th/9711165.

\bibitem{HoChu}
C.~Chu and P.~Ho,
``Constrained quantization of open string in background B field and
                  noncommutative D-brane,''
hep-th/9906192.

\bibitem{Schomerus}
V.~Schomerus,
``D-branes and deformation quantization,''
JHEP {\bf 06}, 030 (1999)
hep-th/9903205.

\bibitem{SW}
N.~Seiberg and E.~Witten,
``String Theory and Noncommutative Geometry,''
hep-th/9908142.

\bibitem{WittenNoncomm}
E.~Witten,
``Noncommutative Geometry And String Field Theory,''
Nucl.\ Phys.\ {\bf B268}, 253 (1986).

\bibitem{BS}
D.~Bigatti and L.~Susskind,
``Magnetic fields, branes and noncommutative geometry,''
hep-th/9908056.
}

\end{document}